\documentclass[aps,prd,twocolumn,superscriptaddress,showpacs,bibnotes]{revtex4}

\usepackage{graphicx}
\usepackage{dcolumn}
\usepackage{bm}
\usepackage{epstopdf}
\usepackage{mathtools}
\usepackage{natbib}
\usepackage{captcont}
\usepackage{xcolor}
\usepackage{ulem}
\usepackage{supertabular}

\begin{document}
\title{The Macronova in GRB 050709 and the GRB/macronova connection}
\affiliation{Key Laboratory of Dark Matter and Space Astronomy, Purple Mountain Observatory, Chinese Academy of Sciences, Nanjing 210008, China }
\affiliation{Racah Institute of Physics, The Hebrew University, Jerusalem, 91904, Israel}
\affiliation{University of Chinese Academy of Sciences, Yuquan Road 19, Beijing, 100049, China}
\affiliation{National Astronomical Observatory of Japan, Mitaka, Tokyo 181-8588, Japan}
\affiliation{INAF/Brera Astronomical Observatory, via Bianchi 46, I-23807 Merate (LC), Italy}
\author{Zhi-Ping Jin}
\affiliation{Key Laboratory of Dark Matter and Space Astronomy, Purple Mountain Observatory, Chinese Academy of Sciences, Nanjing 210008, China }
\author{Kenta Hotokezaka}
\affiliation{Racah Institute of Physics, The Hebrew University, Jerusalem, 91904, Israel}
\author{Xiang Li}
\affiliation{Key Laboratory of Dark Matter and Space Astronomy, Purple Mountain Observatory, Chinese Academy of Sciences, Nanjing 210008, China }
\affiliation{University of Chinese Academy of Sciences, Yuquan Road 19, Beijing, 100049, China}
\author{Masaomi Tanaka}
\affiliation{National Astronomical Observatory of Japan, Mitaka, Tokyo 181-8588, Japan}
\author{Paolo D'Avanzo}
\affiliation{INAF/Brera Astronomical Observatory, via Bianchi 46, I-23807 Merate (LC), Italy}
\author{Yi-Zhong Fan$^\ast$}
\affiliation{Key Laboratory of Dark Matter and Space Astronomy, Purple Mountain Observatory, Chinese Academy of Sciences, Nanjing 210008, China }
\affiliation{Collaborative Innovation Center of Modern Astronomy and Space Exploration, \\
Nanjing University, Nanjing, 210046, China}
\author{Stefano Covino$^\ast$}
\affiliation{INAF/Brera Astronomical Observatory, via Bianchi 46, I-23807 Merate (LC), Italy}
\author{Da-Ming Wei}
\affiliation{Key Laboratory of Dark Matter and Space Astronomy, Purple Mountain Observatory, Chinese Academy of Sciences, Nanjing 210008, China }
 \author{Tsvi Piran$^\ast$}
\affiliation{Racah Institute of Physics, The Hebrew University, Jerusalem, 91904, Israel}

\begin{abstract}
GRB 050709 was the first short Gamma-ray Burst (sGRB) with an identified optical afterglow. In this work we report a re-analysis of the publicly available data of this event and the discovery of a  Li-Paczynski macronova/kilonova that dominates the optical/IR signal at  t $>$ 2.5 days.   Such a signal would arise from ~0.05 M$_\odot$ r-process material launched by a compact binary merger. The implied mass ejection supports the suggestion that compact binary mergers  are significant and possibly main sites of heavy r-process nucleosynthesis. We have re-analyzed all afterglow data from nearby short and hybrid GRBs. A statistical study of sGRB/macronova connection reveals that macronova may have taken place in all these GRBs though the fraction as low as  0.18 cannot be ruled out. The identification of two of the three macronova candidates in the I-band implies a more promising detection prospect for ground-based surveys.
\end{abstract}
\maketitle

\section{introduction}
Compact object mergers are  strong sources of gravitational waves
(GW)  and are prime targets for the advanced LIGO/Virgo
detectors \cite{Abadie2010,Aasi2013}.
It has been suggested that short Gamma-ray Bursts (sGRBs) arise from  mergers in which
one of the compact objects is a neutron star \cite{Eichler1989}, {a scenario now favored by a broad range of observations (see e.g.
\cite{Nakar2007,Berger2014})}. In the absence of GW detection,
a clear  signature for the compact-binary origin of a sGRB is a Li-Paczynski macronova/kilonova:
a near-infrared (nIR)/optical transient powered by the radioactive decay of
$r-$process material synthesized in ejecta  launched during the merger \cite{Li1998,Kulkarni2005,Rosswog2005,Metzger2010,Roberts2011,Korobkin12,Barnes2013,Kasen2013,Tanaka2013,Piran2013,Grossman2013,Tanaka2014,Lippuner2015}.

To date,  the evidence of a macronova associated with sGRB 130603B is based on only a single data
point \cite{Tanvir2013,Berger2013}.
The peculiar GRB 060614 was denoted as a  ``hybrid  burst", since its  $T_{90}\approx 102$ s  groups it with long-duration GRBs, while its temporal lag and peak luminosity are within the short-duration GRB subclass \cite{Gehrels2006}. Moreover there is no evidence for an associated supernova emission  \cite{Fynbo2006,Della2006b,Gal-Yam2006b} down to very stringent limits.
The most significant macronova evidence within this afterglow  is  due to a single Hubble Space Telescope (HST) observation at $t\sim 13.6$ days after the burst \cite{Yang2015}.  Further explorations of the afterglow allowed us to derive a  tentative macronova light curve  \cite{Jin2015}.
In search for further evidence for other macronovae we explored the  optical/IR afterglows of all other nearby
short and hybrid GRBs (hGRBs) in which macronova signals could have been detected.
We begin with the study of GRB 050709, the first sGRB with an identified optical afterglow. Previous works have found irregularity in this afterglow and interpreted it as  a jet break \cite{Fox2005} or as an optical flare \cite{Watson2006}. Reanalyzing the previous observations we suggest here  that this irregularity arises due to a macronova component which in fact dominates the afterglow light curve in this burst. We then compare it with other GRBs/macronovae  and explore the implications of these results to the short-GRB/macronova connection.

We have identified a possible macronova in the optical afterglow data of sGRB 050709.  The $I$-band light curve of this macronova candidate is remarkably similar to that of the macronova candidate of hGRB 060614 \cite{Yang2015, Jin2015}, even though
the isotropic-equivalent energy ($E_{\rm \gamma, iso}$) of their prompt emission and the X-ray afterglow light curve
are significantly different.
Examination of the late-time optical$-$nIR data of all nearby short
and hGRBs ($z<0.4$) for which a macronova could have been observed (six in total) revealed  that there are three  events GRBs 050709, 060614, and 130603B
in which a macronova candidate has been detected. The three other events don't show  such a signal but for each one of them there are  concerns that explain this away. The appearance of a macronova candidate in three out of three (or at most six) events suggest that
macronovae are ubiquitous.
This supports strongly the hypothesis that  compact binary mergers that are accompanied by sGRBs, are the prime sites of heavy r-process nucleosynthesis. The identifications of two of those macronova candidates in the $I$-band
suggest that macronova could be more easily detected in GW follow-up searches, even without a GRB trigger.

\section{Results}

\subsection{A macronova signal associated with GRB 050709}

GRB 050709 was detected by the NASA's High Energy Transient Explorer (HETE-2)
and was localized by the HETE-2's Soft X-ray Camera \cite{Villasenor2005}.
Its prompt emission consisted of a hard spike ($\sim 0.5$ s) and an extended X-ray
emission lasting $\sim 130$ s \cite{Villasenor2005}.
The accurate localization led to follow-up observations allowed to identify
the first  optical afterglow of a short gamma-ray burst
\cite{Hjorth2005,Fox2005,Covino2005}. About 1.5 days after the trigger of sGRB 050709,  Hjorth et al. \cite{Hjorth2005} observed it  with the {\it Danish} 1.54m telescope. They reported two  $R$-band detections. Fox et al. \cite{Fox2005} obtained four HST exposures in the  F814W-band.    The HST observed the site of sGRB 050709 one year later in the same band and didn't detect any signal. Covino et al. \cite{Covino2005} observed the source with the Very Large Telescope (VLT) in $V/R/I$ bands and detected the afterglow in $V$ and $R$ bands simultaneously on Jul. 12.4 UT. The optical afterglow that was localized with sub-arcsecond accuracy was in the outskirts of an irregular, late-type
galaxy at a redshift of $0.16$ \cite{Hjorth2005,Fox2005}. The host's star-formation
rate,  $\sim 0.2~{M_\odot~{\rm yr^{-1}}}$,  is much higher than that of the hosts of the two other sGRBs detected at the time, i.e. 050509B and 050724,  and it renders GRB 050709 to be the first sGRB occurring in a star forming ``low-luminosity" galaxy \cite{Hjorth2005,Covino2005}.
The X-ray afterglow observations of sGRB 050709 are scarce. At $t> 200$ s there are only two significant detections by Chandra (including an X-ray flare at $t\sim 16$ days).  Two other  {\it Swift} ($t\sim 1.6$ days) and Chandra ($t\sim 16.1$ days) data points  have a significance of
$\sim 2\sigma$ \cite{Fox2005}. No radio afterglow emission has been detected \cite{Fox2005}.

Already in 2005, Fox et al. \cite{Fox2005} noted that the early HST optical/IR data declined as  $t^{-1.25 \pm 0.09}$ and then it dropped as $t^{-2.83 \pm 0.39} $ between 10 and 20 days. They  suggested that this arose due to a  jet break.
This interpretation was valid for the HST data set available at that time.  Later,
Watson et al. \cite{Watson2006} combined  the optical/near-infrared (nIR)  data from the Danish 1.54m telescope, VLT and HST,  and showed that the decline is much faster: a single power law of $t^{-1.73 \pm 0.04}$. A single HST data point at $t\sim 9.8$ days was significantly above this line and this was interpreted as a flare powered by a central engine activity.
Following a re-analysis of all publicly available data we show that the light curve is chromatic and this rules out an afterglow scenario (e.g. a jet break). We find a strong evidence for the presence of a new emission component besides the regular forward shock emission and that this component is strong not just at $t\sim 9.8$ days but also at earlier times. We compare the light curve to the predictions of macronova estimates and we  suggest that this nIR excess lasting $\sim 10$ days indicates a macronova emission.

We have analyzed  all publicly-available optical/nIR data of the afterglow of sGRB 050709 (see the Methods for the details).  Results of our re-analysis are generally consistent with those reported in the literature \cite{Covino2005,Fox2005}.
For the VLT data at $\sim 2.5$ days, we confirm the detection in $R$ and $V$ bands.
However, while previous analysis of this data yielded only  an upper limit of $23.25$ mag \cite{Covino2005} our reanalysis of the VLT $I$-band at $t\sim 2.5$ days resulted in a detection with  a Vega magnitude of $24.1\pm 0.2$
(see the Methods for a detailed discussion of this analysis). This $I$-band observation was almost coincidental with the $R$ and $V$ observations with which we can reliably estimate the energy distribution of the spectrum (SED).  We have also found an unpublished  Gemini optical observation giving a tight $3\sigma$ upper limit of $25.4$ mag on the $R$-band  flux  at $t\sim 6.6$ days after the burst.

\begin{figure}
\begin{center}
\includegraphics[width=0.5\textwidth]{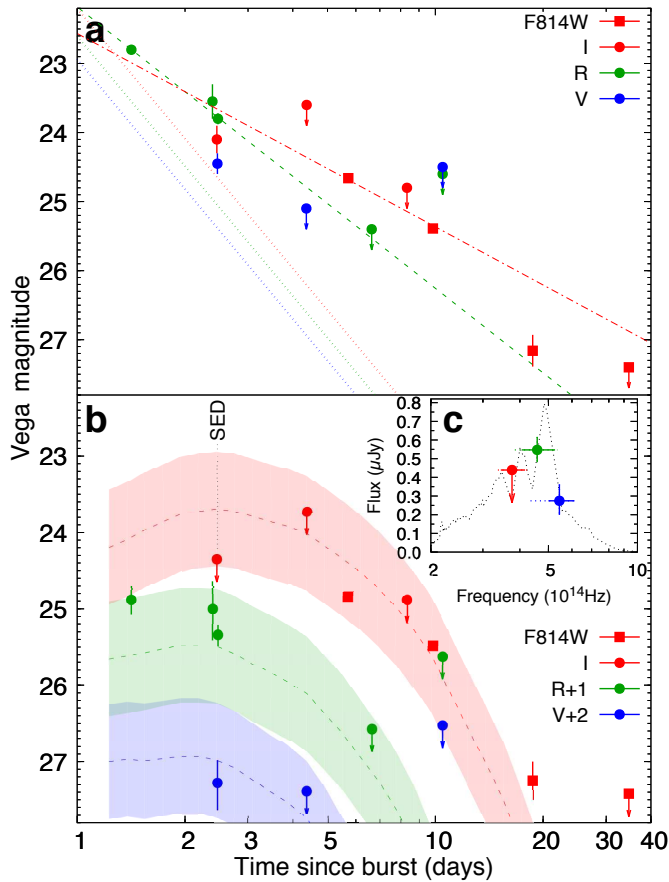}
\end{center}
\caption{{\bf The optical observations of sGRB 050709 (a) and
 a comparison of the data with a  theoretical macronova light curve (b)}. (a): The fits to the  $R$-band emission (green dashed line)
and to the $I$-band observations from the VLT $I$-band data as well as the first two HST F814W-band data points (red dash-dotted line) yield the declines of $t^{-1.63\pm 0.16}$ and $t^{-1.12\pm 0.09}$, respectively.
{The dotted lines represent the ``suggested"-afterglow emission lightcurves of the GRB outflow after the jet break (i.e., $t^{-2.5}$ for the energy distribution index of the shock-accelerated electrons $p\sim 2.5$).}
(b): Shown are the residuals of the optical emission after the subtraction of a suggested fast-declining forward shock afterglow after $t=1.4$ days (dotted lines  in the upper panel). The simulated $I/R/V$-band macronova light curves \cite{Tanaka2014} are for the ejecta from a black hole$-$neutron star merger, corresponding to an ejection mass of $M_{\rm ej}\sim 0.05~M_\odot$ and a velocity of $V_{\rm ej}\sim 0.2c$.  An uncertainty of $\sim 0.75$ mag (the shaded region) has been adopted following Hotokezaka et al \cite{Hotokezaka2013}. (c): The SED of the macronova signal of sGRB 050709 measured by VLT on July 12, 2005 compared with a possible Iron line-like spectral structure adopted from Kasen et al. \cite{Kasen2013}. Note that all errors are $1\sigma$ statistical errors and the upper limits are at the $3\sigma$ confidence level.
}
\label{fig:LC}
\end{figure}

The Fig.\ref{fig:LC}(a) depicts all the available optical/nIR data.
The $R$-band emission decreases  as $t^{-1.63\pm0.16}$. This is consistent with the $V$-band data and with the overall fit of Watson et al. \cite{Watson2006}. Note that the new  $R$ band upper limit (at $t\sim 6.6$ days) is consistent with the Watson et al. \cite{Watson2006}'s fit. On the other hand the $I$-band emission decreases much slower, as  of $t^{-1.12\pm0.09}$, and this is consistent with the Fox et al. \cite{Fox2005}'s analysis of the early HST data alone. The standard afterglow model, implies an achromatic decay and hence the different  behavior in the $R$ and $I$ bands over a long timescale of $\sim 10$ days  is inconsistent with an afterglow model \cite{Piran2004}.
In fact attempt to fit all the $I$ and $R$ band observations to a single achromatic broken power fails, with the best $\chi$-square p.d.f.
obtained is of order 10 (ignoring the HST data point at $t\sim 18.7$ days does not solve the problem). This strongly suggests  an additional component.

Further information is obtained from the puzzling spectrum at $t\sim 2.5$ days. Here, the $R$ band flux is significantly larger than the $I$ band flux. This again is inconsistent with a standard afterglow model and it  suggests that an additional component dominates already at this stage. Namely, any afterglow emission is subdominant already at $t\sim 2.5$ days. This can happen if there was an early jet break at $t \lesssim 1.4$ days, in which case the afterglow would have declines from its observed value at $t=1.4$ days as $t^{-p}$ , with $p > 2$. Such a decline (with $p =2.5$, which is consistent with the X-ray spectrum)  is also shown in Fig.\ref{fig:LC}.
Indeed, for $p=2.5$ and the cooling frequency $\nu_{\rm c}\sim 2.5\times 10^{16}$ Hz at $t\sim 2.5$ days, the extrapolation of the Chandra X-ray emission into optical bands yields emission flux lower than the VLT data, consistent with  the presence of a macronova emission component.
 Both the required fast decline rate as well as the jet break time are consistent with that observed in some other sGRBs and in particular in sGRB 130603B and hGRB 060614 \cite{Tanvir2013,Berger2013,Xu2009}, two events displaying macronova signals. Remarkably, even without the VLT I-band data, Watson et al. \cite{Watson2006} already noticed
that the decline is rather steep suggesting a post jet break afterglow and that
 at $t\sim 10$ days the HST F814W-band emission was in excess of the regular forward shock afterglow emission. With the new data points the evidence for a macronova signal is much stronger. Remarkably, this $I$/F814W macronova signal (see Fig.\ref{fig:nIR}, where the suggested-afterglow component has been subtracted) is very similar to that identified in hGRB 060614 \cite{Jin2015}.

In Fig.\ref{fig:LC}(b) we compared the observed lightcurves with the predictions of a macronova model. Shown are the residual of the optical emission after the subtraction of a suggested forward shock afterglow  with a fast declining emission after $t=1.4$ days and the theoretical lightcurves of  a macronova following a black hole$-$neutron star merger \cite{Tanaka2014} with  $M_{\rm ej} \sim 0.05~M_\odot$ and $v_{\rm ej}\sim 0.2c$, where $c$ is the speed of light, $M_{\rm ej}$ and $v_{\rm ej}$ are the ejecta mass and velocity, respectively. This is comparable but slightly smaller than the parameters used  for fitting the $I$-band excess observed  in the afterglow of GRB 060614 \cite{Yang2015}. Such a large amount of $r$-process material is consistent with a black-hole neutron star mergers \cite{Foucart2014,Just2015,Kyutoku2015,Kiuchi2015} and it also supports  the hypothesis that compact object mergers are prime sites of significant production of $r-$process elements \cite{Lattimer1977,Eichler1989,Piran14,Hotokezaka2015,
Shen2015,Voort2015,Ishimaru2015,Wehmeyer2015}. The black-hole neutron star merger scenario also has a significant implication on the prospect of establishing the GRB/GW connection in the advanced LIGO/Virgo era \cite{LiX2016}.

The weak $I$-band emission at $t\sim 2.5$ days together with the almost simultaneous $R$ and $V$ observations, implies a puzzling broad line-like structure (see Fig.\ref{fig:LC}(c) for the afterglow-subtracted SED).
A speculative interpretation is that this signal is due to a disk wind driven macronova.
A strong line feature can be produced by a macronova dominated by Iron \cite{Kasen2013}.
Such an Iron-group dominated macronova may arise from an accretion disk wind~\cite{Metzger2009}
in which the heavier r-process elements are depleted because strong neutrino irradiation from a remnant neutron star or the accretion torus
can increase the electron fraction of the disk material. An interesting possibility is that the sub-relativistic neutron-rich ejecta from the compact object mergers may have a heavier or lighter composition in different directions and  the resulting signal may be a combination of macronovae resulting from those (e.g.  \cite{Metzger2014,Kasen2015}).  A telescope of the E-ELT (European Extremely Large Telescope) class will be able to carry out spectroscopy of these faint signals allowing a better understanding of the phenomena.

Before concluding we note that if
we do not rely on the re-analysis of the data, and adopt the afterglow interpretation of Watson et al. \cite{Watson2006},
even in this case there is an $I$ band excess at $9.8$ days. The most natural explanation for this excess is also a macronova and the physical parameters are similar to that adopted in the modeling of Fig.\ref{fig:LC}.

\subsection{Macronvovae are ubiquitous in afterglows of short and hybrid GRBs}\label{sec:GRBs}

Following the tentative discovery of a third macronova signal we have re-examined all nearby sGRBs and hGRBs to search for  possible macronova signals. Usually the macronova optical spectrum is expected to be soft, therefore ground-based deep $I$-band observations (ground-based $J/H/K$-band observations usually are not deep enough) as well as  HST nIR observations are essential. The macronova candidates emerged in the sGRB 130603B, hGRB 060614 and sGRB 050709 lightcurves 1-2 weeks after the GRB triggers. At earlier times the forward shock afterglow emission outshines  the macronova component while at late times the macronova emission also faded away.
Hence we need deep $I$-band or  near-infrared HST observations in the time interval of $\sim 5-15$ days. Theoretical predictions for macronovae
vary significantly depending on the ejecta mass  $M_{\rm ej}$, the velocity  $V_{\rm ej}$, the composition, the merger types, and
different observing angles (see e.g. Fig.10 of \cite{Tanaka2013} and Fig.9 of \cite{Tanaka2014} for illustration).
For a reference we note that the observed signatures were  $\sim 24.5$ Vega mag at about 9 days in F160W ($H$) band for sGRB 130603B at redshift 0.356, $\sim 25$ Vega mag at about 13.5 days in F814W ($I$) band for hGRB 060614 at redshift 0.125 and $\sim 25$ mag (Vega) at about 10 days in F814W($I$)  band for sGRB 050709 at redshift 0.16.

\begin{figure}
\begin{center}
\includegraphics[width=0.5\textwidth]{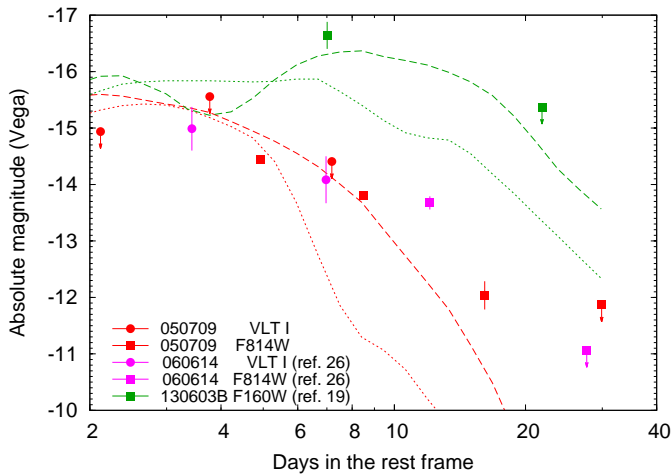}
\end{center}
\caption{{\bf Comparison of the lightcurves of macronova candidates and theoretical models.} Absolute Vega magnitudes versus rest frame time of the macronova candidates in sGRB 050709, hGRB 060614 \cite{Jin2015} and sGRB 130603B \cite{Tanvir2013}.
The red dashed line is the same as the dynamical ejecta macronova model $I$-band emission presented in Fig.\ref{fig:LC} (the green dashed line represents the $H$-band emission) while the red dotted line is the disk-wind ejecta macronova model $I$-band emission light curve \cite{Tanaka2016b} for $M_{\rm ej}=0.03~M_\odot$ and $V_{\rm ej}=0.07c$ (the green dotted line represents the $H$-band emission). Note that all errors are $1\sigma$ statistical errors and the upper limits are at the $3\sigma$ confidence level.
}
\label{fig:nIR}
\end{figure}

We focus on  {\it Swift} and HETE-2 sGRBs and hGRBs at redshifts $z\leq 0.4$ since  HST observations  needed
for such observations at higher redshifts are  scarce \cite{Berger2014,Fong2015}.
The initial ``low redshift" sample consists of  sGRBs   050509B,  050709,  050724,  060502B,  061201,   071227,  080905A,  130603B,  140903A,  and 150101B and  hGRBs 060505 and 060614 \cite{Berger2014,Fong2015}.
Unfortunately most of these GRBs are not suitable and have to be excluded from the ``macronova  candidates" sample.
There were no observations within the macronova phase for  sGRBs 050724, 060502B, 071227, 080905A and 140903A.
No such observations were published yet for sGRB 150101B.
The $I$/nIR observation information of the remaining events, sGRBs 050509B, 050709, 061201 and 130603B and hGRBs 060505 and 060614
are summarized in Supplementary Table 1 (see the Supplementary Information).
Three events, sGRBs 050509B, 061201 and hGRB 060505 are potentially interesting but each one has its own caveat. The suggested host galaxy of sGRB 050509B is very bright and no optical counterpart had been detected.  Hence the upper limits on the ``underlying" afterglow and macronova emission sensitively depend on the unknown location within the host galaxy (see also \cite{Hjorth2005}). The redshift of  GRB 061201 is not secure \cite{DAvanzo2014} and it is possible that it was not sufficiently nearby.
Using the hardness and prompt duration distribution Bromberg et al.
\cite{Bromberg2013} estimate that hGRB 060505 has a $97^{+2}_{-22}\%$ probability of being a Collapsar (see also the argument based on the location of the burst within a bright star forming region \cite{Ofek2007b} and  host galaxy observations \cite{Thone2014}).

\begin{figure}
\begin{center}
\includegraphics[width=0.5\textwidth]{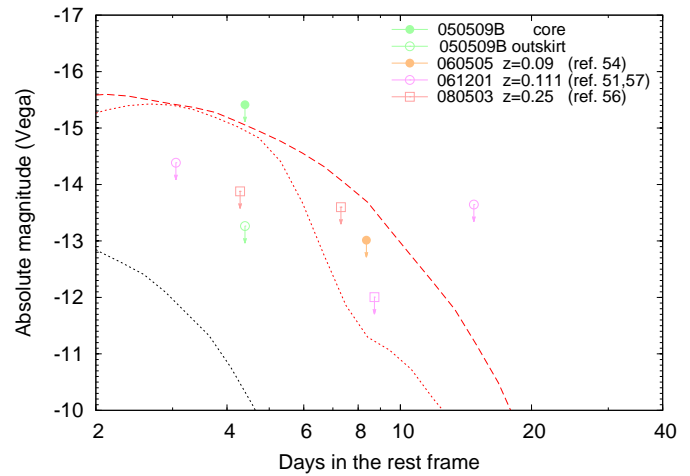}
\end{center}
\caption{{\bf Comparison of the limits of macronova in some sGRBs and theoretical models.} Absolute Vega magnitudes versus rest frame time of the I-band/F814W-band observations of sGRB 050509B,
hGRB 060505 \cite{Ofek2007b} and sGRB 061201 \cite{Stratta2007b,Fong2015}. The HST F814W  {$3\sigma$} upper limits of GRB 080503 \cite{Perley2009} are also shown for an assumed redshift of $z=0.25$, following Kasen et al. \citep{Kasen2015}.  Note that the Gemini $i$-band  {$3\sigma$} upper limit of sGRB 060505 was re-analyzed in this work. The red dashed line is the dynamical ejecta macronova model $I$-band emission while the red solid line is the disk-wind ejecta macronova model $I$-band emission light curve, where the same model parameters in Fig.2 are chosen. The black dotted line represents the macronova I-band emission expected for a double neutron star merger \citep{Tanaka2013} with $M_{\rm ej}=0.01~M_\odot$ and $V_{\rm ej}\sim 0.1c$, implying that the  {$3\sigma$} upper limits reported in sGRB 050509B,
hGRB 060505 \cite{Ofek2007b} and sGRB 061201 are not deep enough to exclude the compact object merger origin.
}
\label{fig:nIR-UL}
\end{figure}

Therefore, in total there are just three or at most six events
that are sufficiently nearby and have sufficient data for a macronova identification. In the three of those (sGRB 050709, hGRB 060614 and sGRB 130603B) there are macronovae signatures (see Fig.\ref{fig:nIR}).  In the three others potentially interesting events (sGRB 050509B, hGRB 060505 and sGRB 061201) there are only  upper limits (see Fig.\ref{fig:nIR-UL})  but it is possible that none of them is sufficiently binding.
In the most ``optimistic" case there are 3 macronovae in a sample consisting of just three events and  the 95\%  confidence interval of the probability of a macronova taking place in a sGRB/hGRB is $(0.47,~1)$. While in the most ``pessimistic" case (i.e., there are 3 macronovae in a sample consisting of six GRBs) the 95\%  confidence interval for the probability is $(0.18,~0.82)$. Therefore the detection prospect of macronovae in merger-powered GRBs are indeed encouraging though the fraction as low as  $\sim 0.18$ cannot be ruled out.

Within this context it is interesting to mention  also GRB 080503. It is not in our sample as its redshift is unknown \cite{Perley2009}. Though no $I$-band/F814W-band or redder emission had been measured (see Fig.\ref{fig:nIR-UL}, where the upper limits on the infrared luminosity are for a redshift $z\sim 0.25$, as assumed by Kasen et al. \cite{Kasen2015}), in optical bands the afterglow was detected in the time interval of $\sim 1.08-5.36$ days after the GRB trigger. The emission is quite blue, which is at odds with the dynamical ejecta macronova model but may be consistent with the disk-wind macronova model \cite{Kasen2015}. The potential challenge for this model is the non-identification of a nearby host galaxy as close as $z\sim 0.25$ in the deep HST/WFPC2 observation data of GRB 080503 \cite{Perley2009}.

It is interesting to compare now the observed features of the three macronova candidates.
As far as the prompt emission is  concerned, GRB 050709, a short burst with extended soft X-ray emission, bridges the  gap between the canonical sGRB 130603B and the hGRB 060614 (see Table 1). The isotropic-equivalent prompt emission energy $E_{\rm \gamma,iso}$ of sGRB 050709 is about 30 times smaller than that of hGRB 060614 and sGRB 130603B,
while the macronova emission of sGRB 050709 is similar to that of hGRB 060614 (see Fig.\ref{fig:nIR}).
The high energy transients were powered by
a relativistic jet emerging from the central engine while the macronova emission arises from the $r$-process material ejected during the merger. The similarity between the macronova emission of sGRB 050709 and hGRB 060614  that had a  very different energy release in the prompt phase suggests that the launch processes of the ultra-relativistic outflows and the sub-relativistic outflows are not related.

\begin{tiny}
\begin{table*}
\label{tab-GRBs}
\begin{center}
\title{}Table 1. Physical properties of GRBs/macronovae/afterglows with known redshifts.\\
\begin{tabular}{llllll}
\hline
		&	GRB 050709$^{a}$		&	GRB 060614$^{b}$			&	GRB 130603B$^{c}$	 \\
\hline
$E_{\rm \gamma, iso}$ ($10^{51}$ erg)	&	0.069 	&	2.5	&	2.1 	\\
$z$	&	0.16	&	0.125	&	0.356			 \\
Duration$^{d}$ (s)	&	0.5 (+130) 	&	5 (+97)	&	0.18		 \\
Classification 	&	sGRB $+$ extended X-rays &	hGRB	&	sGRB		 \\
Identifying macronova	&	in $I$/F814W	&	in $I$/F814W	&	in F160W	\\
Macronova peak luminosity    &	 $\sim 10^{41}~{\rm erg~s^{-1}}$ ($I$)	&	 $\sim 10^{41}~{\rm erg~s^{-1}}$ ($I$)	&	$\sim 10^{41}~{\rm erg~s^{-1}}$ (F160W)                   \\
$M_{\rm ej}$$^{e}$ 	&	$\sim 0.05~M_\odot$	&	$\sim 0.1~M_\odot$	&	$\sim 0.03~M_\odot$	\\
${\cal R}_{\rm MN/X}$$^{f}$ &         $\sim 1$         &               $\sim 0.1$               &             $\sim 0.4$              \\
\hline
\end{tabular}
\end{center}
Note: a. Villasenor et al.\cite{Villasenor2005} and this work; b. Gehrels et al. \cite{Gehrels2006}, Yang et al. \cite{Yang2015} and Jin et al. \cite{Jin2015}; c. Tanvir et al. \cite{Tanvir2013}, Berger \cite{Berger2013} and Hotokezaka et al. \cite{Hotokezaka2013}; d. The durations include that of the hard spike and the ``extended emission" (in the bracket); e. The $M_{\rm ej}$ is estimated from the dynamical ejecta model and the value can change by a factor of a few due to uncertainties in the opacity, nuclear heating, and ejecta morphology; f. ${\cal R}_{\rm MN/X}$ denotes the ratio between the macronova ``peak" luminosity and the simultaneous X-ray luminosity.
\end{table*}
\end{tiny}

At $t\sim 16$ days after the trigger of sGRB 050709 there was an X-ray flare \cite{Fox2005}. While at $t>1.4$ days after the trigger of hGRB 060614, the X-ray afterglow is well behaved \cite{Xu2009}. At $t\gtrsim 1$ days after the trigger of GRB 130603B, the X-ray emission became  flattened \cite{Fong2015}. The ratio between the macronova and X-ray radiation luminosities at the peak time of the macronova emission (i.e., ${\cal R}_{\rm MN/X}$) varies from burst to burst by up to a factor of 10. For sGRB 050709, hGRB 060614 and  sGRB 130603B, the ${\cal R}_{\rm MN/X}$ are $\sim (1, ~0.1, ~0.4)$, respectively, which could shed some lights on the physical origin (see below).

A remarkable feature shown in Fig.\ref{fig:nIR} is the comparable peak luminosities of the different  macronovae (i.e., $\sim 10^{41}~{\rm erg~s^{-1}}$). However, the macronovae associated with sGRB 050709 and hGRB 060614 were mainly identified in $I$/F814W-band, which are ``bluer" than the F160W-band macronova component of sGRB 130603B. As in none of the cases we have a complete spectrum it is not clear if there was a real difference in the spectra.

\section{Discussion}

The possible identification of three macronova candidates in a small sample containing just three or at most
six events that are suitable for the search indicates that
macronovae are common in sGRBs and hGRBs. Given the paucity
of data for other events, macronovae could possibly arise in all sGRBs though a macronova fraction as low as $\sim 0.18$ can not be ruled out.
A common feature of the macronova candidates is
 that the peak luminosity of macronovae is $\sim 10^{40}-10^{41}~{\rm erg~s^{-1}}$
in the optical to infrared bands with a timescale of one week.
In the compact binary merger scenario of sGRBs, this  can arise from
dynamical ejecta with heavy $r$-process elements, or lanthanide-free wind, or central engine activity.
Here we discuss implications to each model.

The $I$-band light curve arising from dynamical ejecta with a mass of $0.05~M_{\odot}$ and
an average velocity of $0.2c$~(see the black-hole
neutron star merger model H4Q3a75 in \cite{Tanaka2014} and also  \cite{Kawaguchi2016})
is shown in Fig.1 and Fig.2 as an example.
Because of a fast expansion velocity and the large opacity of $\approx 10~{\rm cm^{2}~g^{-1}}$,
the temperature is already  low around the peak time and
most of the  photons are radiated in the near infrared $J$, $H$, and $K$-bands.
The  luminosity in the $I$-band is smaller than that  in the $H$-band
by a factor of $3-10$ at $1$ week after merger.
This model can reproduce the observed $I$-band data of sGRB 050709 and the hGRB 060614
and the $H$-band data of sGRB 130603B with
$\sim (0.05~M_{\odot},~0.1~M_{\odot},~0.03M_{\odot})$,  respectively.
The  massive ejecta with $\gtrsim 0.05M_{\odot}$ suggests that the progenitor of sGRB 050709 is a black-hole neutron star merger \cite{Foucart2014,Just2015,Kyutoku2015,Kiuchi2015}.
However we should note that this estimate can change by a factor of a few due to uncertainties in
the opacity, nuclear heating, and ejecta morphology.
The upper limit in the $I$-band at $3$ days of sGRB 061201 constrains
the maximally allowed mass of dynamical ejecta as $\approx 0.02M_{\odot}$
if the redshift of $0.111$ is correct.  The upper limits in the $I$ band of hGRB 060505 are
both consistent with almost $0.05 M_{\odot}$ and can be even higher if absorption at the host
galaxy  was significant \cite{Ofek2007b}.
Interestingly \cite{Ofek2007b} was the first to search for a macronova signature in the afterglow light curve of this burst.

The absence of lanthanides in a wind reduces the opacity. The resulting macronova
has a brighter and bluer peak luminosity on a shorter timescale \cite{Kasen2013,Tanaka2013,Metzger2014,Perego2014,Kasen2015}.
Figure 2 shows the $I$-band light curve arising from a lanthanide-free wind
with a mass of $0.03M_{\odot}$ and an average velocity of $0.07c$,
where elements with atomic numbers of $31-54$ are included
(see the wind model in \cite{Tanaka2013} and also \cite{Kasen2015}).
This model can reproduce the $I$-band data of GRB 050709 and 060614
at early times ($t<5$ days). However, the light curve at late times
is faint compared to the data.  While increasing the wind mass raises
the late $I$-band luminosity, such a model is too bright to be compatible
with the early $I$-band data~($t<5$~days).
The $I$-band upper limit at $3$~days of GRB 061201 indicates
the mass of a lanthanide-free wind of $\lesssim 0.01M_{\odot}$. The upper limits on
the afterglow of hGRB 060505 were taken much later after the bursts and as such the implied limits on
the wind ejecta are weak.

The central engine can also power a macronova.  Here we focus  on
the X-ray powered macronova  model \cite{Kisaka2015} since this model is
testable with the observed X-ray and optical data.
In this model, X-ray photons emitted by the central engine are absorbed by the ejecta and
re-emitted in the optical-infrared bands.
Note that $r$-process material with a mass of $\gtrsim10^{-3}M_{\odot}$
is required in order to keep the ejecta optically thick to optical photons
until one week after the merger. While the spectrum and light curve of this emission
are unclear, a relation of $L_{\rm IR}\approx 0.1L_{X}$ (i.e., ${\cal R}_{\rm MN/X}\approx 0.1$)
 is expected in this scenario. As summarized in Table 1, ${\cal R}_{\rm MN/X}$ varies among the events.
In particular, for GRB 050709, it is difficult to explain the macronova luminosity with
${\cal R}_{\rm MN/X}\approx 1$.  However, the flare activity in X-ray at late times
may provide enough energy to produce the $I$-band emission.
Better data in both X-ray and optical-IR at late times
are needed to further test the X-ray powered macronova model.

The  comprehensive examination of the near-infrared data of current nearby sGRBs and hGRBs  yielded
in total three or at most six events suitable for macronova searches. The successful identification of three candidates in such a limited sample demonstrates that  macronovae arise in most if not all compact object merger events that produce GRBs.
A comparison of the above three scenarios favors the by now ``standard" dynamical ejecta that is enriched by
$r$-process elements \cite{Korobkin12,Wanajo2014,Just2015}.
The massive $r$-process material ejecta inferred
in each one of these  events strongly suggest that compact object mergers are the significant or even prime sites of producing heavy  $r$-process elements \cite{Lattimer1977,Eichler1989,Piran14,Hotokezaka2015,
Shen2015,Voort2015,Ishimaru2015,Wehmeyer2015}.

These results have important implications on the future of macronovae and
GW electromagnetic counterparts observations (see e.g., \cite{Nissanke2013,Berger2015} for search strategies). Among the three macronova candidates, two were identified in $I$-band (there was also evidence for emission in R-band emission, too). Ground-based telescopes are much more sensitive in  $I$-band  than in $J/H/K$-bands. If the mergers powering sGRB 050709 and hGRB 060614 took place at luminosity distances of $\sim 200$ Mpc (the horizon of advanced LIGO/Virgo network for double neutron star mergers) or $\sim 350$ Mpc (the horizon of advanced LIGO/Virgo network for a neutron star merger with a $\sim 6~M_\odot$ black hole), the corresponding peak I-band emission is expected to be as bright as $\sim 21-22$th magnitude or $\sim 22-23$th magnitude, respectively. Such events  are marginally detectable by new and upcoming transient surveys such as the ESO VLT Survey Telescope (VST, https://www.eso.org/sci/facilities/paranal/telescopes/vst.html, see Abbott et al.\cite{Abbott2016b}) and
the Zwicky Transient Facility (ZTF) that is expected to have first light in 2017 (http://www.ptf.caltech.edu/ztf). The Large Synoptic Survey Telescope (LSST; \cite{LSST2009}) with a $9.6~{\rm deg^{2}}$ field of view that can image about $10^{4}~{\rm deg^{2}}$ of the sky in three clear nights down to limiting magnitude  of $i \sim 23.5$ (Vega system), in principle, could easily identify such signals.\\

\section{Methods}\label{sec:datareduction}
\subsection{Optical and infrared data reduction}
\begin{figure*}
\begin{center}
\includegraphics[width=0.32\textwidth]{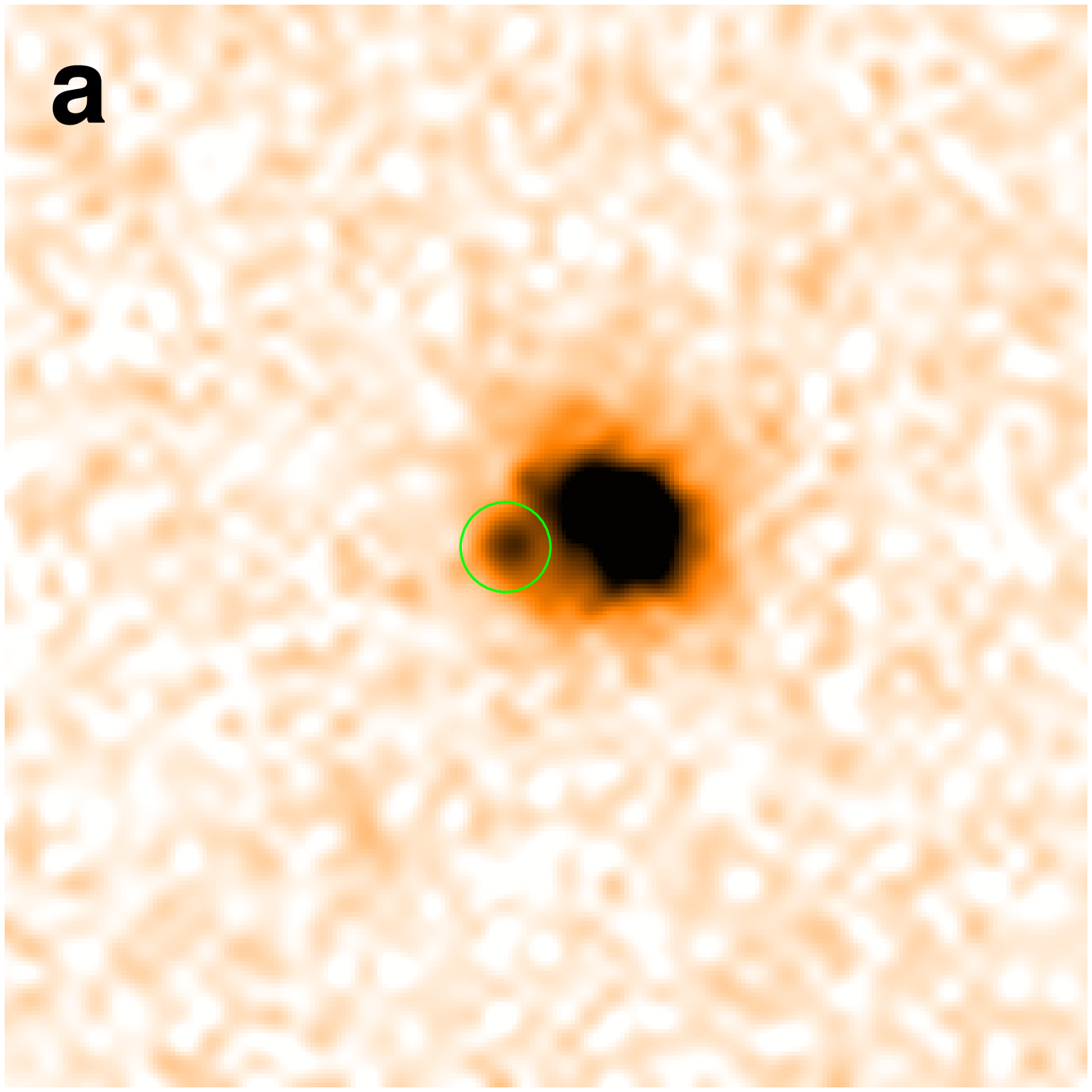}
\includegraphics[width=0.32\textwidth]{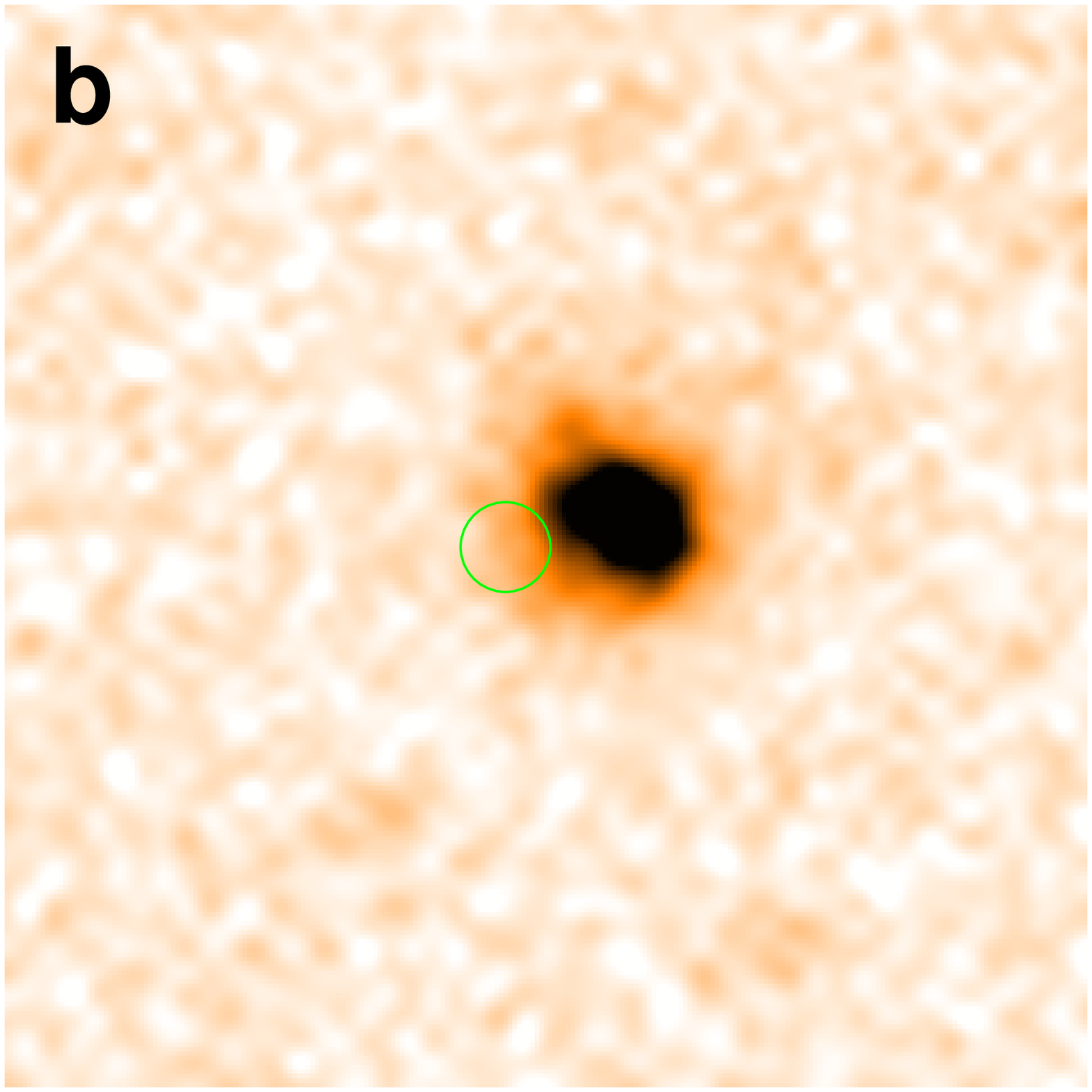}
\includegraphics[width=0.32\textwidth]{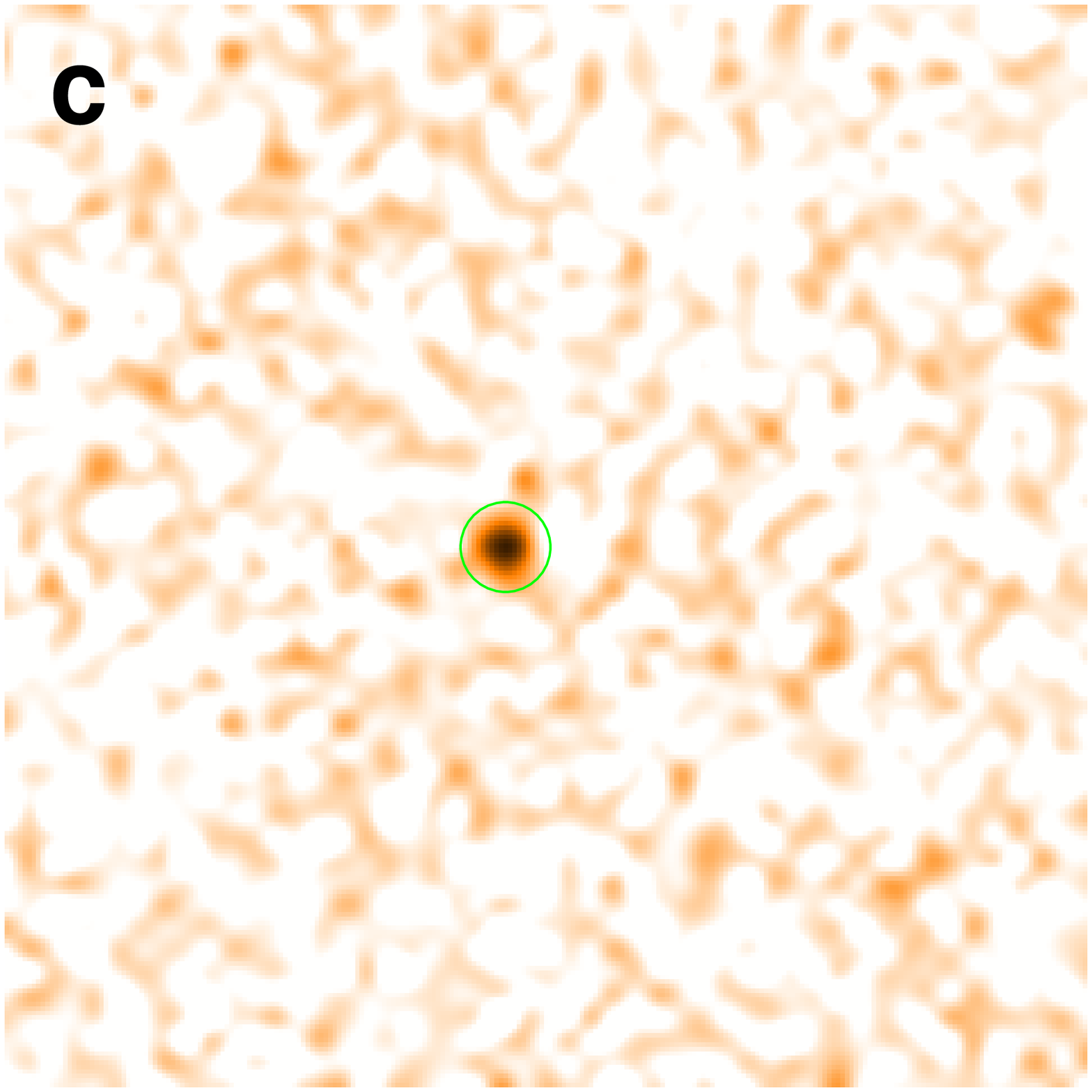}
\end{center}
\caption{{\bf The VLT I band images of the afterglow of GRB 050709.} The data were taken on July 12 2005 (a) and July 30 2005 (b),  and the signal resulted in the image substraction (c). The afterglow position has been circled and the afterglow emission is clearly visible on July 12, 2005. The images are magnified only for demonstration. }
\label{fig:$I$-band}
\end{figure*}

The VLT imaging data of GRB\,050709 are publicly available in ESO Science Archive Facility (http://archive.eso.org).
We reduce the raw data following the standard procedures in IRAF (http://iraf.noao.edu), including bias subtraction, flat fielding and image combination. Observations made with the same filter at different epochs are firstly aligned to the last epoch (reference frame), using the {\it imalign} tool in IRAF. The task {\it ficonv} in software package FITSH (http://fitsh.szofi.net) is used to convolve the reference to match the profile and brightness of objects in earlier frames.
For each earlier frame, the reference frames is convolved to and subtracted. In this method, the object profile and zero point of the subtracted image are the same as the image that has been subtracted.
Finally the aperture photometry is applied to the residual images and find the instrumental magnitudes of the afterglow.
Photometric errors are estimated from the photon noise and the sky variance to $1\sigma$ confidence level.
The 3\,$\sigma$ of the background RMS of the residual images is taken as the limiting magnitude.
Eight to ten point like objects in the field are used as reference stars for differential photometry.
Finally, standard stars observed on July 12 and 30, 2005 were used for the absolute calibration.
The results are presented in Table \ref{tab:opticalobservations}, consistent with that reported in the literature \cite{Covino2005}. The main novel result is the detection of the $I$-band emission at $t\sim 2.4$ day after the trigger of the GRB (see Fig.\ref{fig:$I$-band}).
Our ``new" detection is mainly benefited from the improvement on the change of the reference frame (i.e., from July 18 FORS1 observation to July 30 FORS2 observation). The advantage is less ``contamination" from emission of the source (about 25.2 Mag versus $>27.2$ Mag, according to the HST observation), the original and reference images are both from the same instrument (FORS2) on VLT.  Hence the signal to noise rate of the source is improved.

\begin{tiny}
\begin{table*}
\label{tab:opticalobservations}
\begin{center}
\title{}Table 2. The optical observations of GRB 050709\\
\begin{tabular}{lllllll} \hline \hline
Time		&	Exposure		&	Filter			&	Magnitude$^{a}$ & Flux & Seeing	& Sky brightness$^{b}$\\
(days)	&	(seconds)		&				&	(Vega)&	($\mu$Jy)		& (arcsec)			&  	\\
\hline
\hline
2.46346	&	$60\times5^{c}$	&	VLT/FORS2/V	&	$24.45\pm0.15$ & $ 0.59\pm0.08$	&	0.76	& 21.79 \\
4.36416	&	$120\times3$	&	VLT/FORS1/V	&	$>25.1$	&	$<0.31$	&	0.89    & 21.55 \\
10.48568	&	$120\times3$	&	VLT/FORS1/V	&	$>24.5$	&	$<0.55$	&	0.73    & 19.49 \\
20.16693	&	$180\times3$	&	VLT/FORS2/V	&	$-$		&	$-$	&	0.66	& 21.46 \\
2.47249	&	$60\times5$	&	VLT/FORS2/R	&	$23.80\pm0.08$ & $ 0.90\pm0.07$ &	0.68	& 21.19 \\
6.64339	&	$300\times4^{d}$	&	Gemini-N/r'(R)	&	$>25.4$  & $<0.20$     &	0.67	& 20.93 \\
10.47943	&	$120\times3$	&	VLT/FORS1/R	&	$>24.6$ &	 $<0.43$	&	0.59	& 19.26 \\
20.17874	&	$180\times15^{c}$	&	VLT/FORS2/R	&	$-$ &	$-$	&	0.61	& 20.97 \\
2.45513	&	$100\times6$	&	VLT/FORS2/I	&	$24.1\pm0.2$ &	$ 0.55\pm0.09$ &	0.65	& 19.85 \\
4.37179	&	$100\times3$	&	VLT/FORS1/I	&	$>23.6$ &		$<0.86$ &	0.79	& 19.51 \\
8.33429	&	$120\times10$	&	VLT/FORS1/I	&	$>24.8$ &		$<0.30$ &	0.42	& 19.40 \\
20.23152	&	$180\times3$	&	VLT/FORS2/I	&	$-$ &		$-$	&	0.61	& 19.73 \\
5.71410	&	6360			&	HST	F814W	&	$24.66\pm0.03^{e}$ & $0.330\pm0.009$ &	& \\
9.84385	&	6360			&	HST	F814W	&	$25.39\pm0.05^{e}$ & $0.169\pm0.008$ &	& \\
18.70269	&	6360			&	HST	F814W	&	$27.16\pm0.23^{e}$ & $0.033\pm0.008$ &	& \\
34.69556	&	6360			&	HST	F814W	&	$>27.4^{e}$	& $<0.026$ & & \\
371.78780	&	7039			&	HST	F814W	&	$-$ &		$-$	&	& \\
1.4166	&	$600\times12$	&	Danish/R		&	$22.80\pm0.07^{f}$ & $2.34\pm0.12$	&	& \\
2.3862	&	$600\times17$	&	Danish/R		&	$23.55\pm0.25^{f}$ & $1.17\pm0.26$	&	& \\
\hline
\end{tabular}
\end{center}
Note:
a. The magnitudes of the extracted optical transient, here magnitude errors are reported in 1$\sigma$ and upper limits are 3$\sigma$. \\
b. In units of Vega Mag arcsec$^{-2}$.\\
c. Some images are not combined.\\
d. The Gemini-N r'-band upper limit has been converted into $R$-band.\\
e. Fox et al. \cite{Fox2005} reported the AB magnitudes, which are larger than the corresponding Vega magnitudes by $0.42$ mag. \\
f. These data are taken from Hjorth et al. \cite{Hjorth2005}.\\
\end{table*}
\end{tiny}

Note that in the direction of the burst, the Galactic extinction is expected to be just $E(B-V) =0.01$ mag \cite{Schlegel1998}.
The optical afterglow of GRB 050709 is superposed on the outskirts of the host galaxy and the extinction is probably very small (i.e., $\leq 0.1$ mag), too  \cite{Fox2005}. Therefore in this work, we ignore the extinction corrections of the optical data. We have also analyzed the Gemini-N $r'$-band data. In total there are two sets of exposures (i.e., $4\times 300$s on Jul 16 and $6\times 200$s on Jul 28). However the second sets of exposures have a high sky brightness ($\approx 19.35~{\rm Vega~Mag~arcsec^{-2}}$) that is not suitable for reference frame in the image subtraction. Therefore, we performed image subtraction between the high-quality Gemini-N (on Jul 16) and VLT (on Jul 30) observations and got an upper limit (see Table 2).

We download the public HST archive data of GRB\,050709 from the Mikulski Archive for Space Telescopes (MAST; http://archive.stsci.edu),
including five observations with ACS in F814W band. The reduced data provided by MAST were used in our analysis. The last observation is taken as the reference and the other images of the same filter are subtracted in order to directly measure fluxes of the afterglow from the residual images. Aperture photometry was carried out for the afterglow in the residual image.
The ACS zeropoints were used for absolute calibration. If the signal of the afterglow is too faint to be a secure detection, an upper limit of 3\,$\sigma$ background RMS is adopted. Our results are summarized in Table 2, nicely in agreement with Fox et al. \cite{Fox2005}.

Danish 1.54m telescope data are not publicly available, and therefore we simply adopt the data reported in Hjorth et al. \cite{Hjorth2005}.
Table 2 is a complete list of the data points used in our analysis.

\section{\it End notes.}
{\it Acknowledgement.} This work was supported in part by the National Basic Research Programme of China (No. 2013CB837000 and No. 2014CB845800), NSFC under grants 11525313 (i.e., Funds for Distinguished Young Scholars), 11361140349 (Joint NSFC-ISF Research Program funded by the National Natural Science Foundation of China and the Israel Science Foundation), 11273063, 11433009, 11103084, and U1231101, the Chinese Academy of Sciences via the Strategic Priority Research Program (Grant No. XDB09000000) and the External Cooperation Program of BIC (No. 114332KYSB20160007), the Israel ISF$-$CNSF grant, the Templeton foundation and the I-Core center for excellence ``Origins" of the ISF. SC has been supported by ASI grant I/004/11/0. \\

{\it Author contribution.} Y.Z.F, Z.P.J, S.C and T.P launched this project. Z.P.J, X.L (from PMO), S.C and P.D (from INAF/OAB) carried out the data analysis/statistics. K.H and T.P (from HU), M.T. (from NAOJ), Y.Z.F, Z.P.J and D.M.W (from PMO) interpreted the data. Y.Z.F, T.P, S.C and K.H prepared the paper and all authors joined the discussion.

{\it Author information.} Correspondence should be addressed to Y. Z.~Fan (e-mail: yzfan@pmo.ac.cn), or S. Covino (e-mail: stefano.covino@brera.inaf.it), or T. Piran (e-mail: tsvi.piran@mail.huji.ac.il).

{\it Competing financial interests.} The authors declare no competing financial interests.

\clearpage

\begin{appendix}


\section*{Supplementary discussion: the near-infrared observations of nearby sGRBs and hGRBs}\label{sec:bursts-observations}
In the second part of Sec. II
we have a sample of 12 nearby sGRBs and hGRBs.
The near-infrared observations of these events are summarized in Supplementary Table 1. Note that for the afterglow emission of GRB 150101B, there is still no formally published paper, yet, and we collect the information from GCN Circulars as well as the websites for 8-10m class telescopes and HST. For other events we collect the data reported in the literature. Below we explain in some detail why we conclude that sGRB 060502B, sGRB 050724, sGRB 071227, sGRB 080905A, sGRB 140903A and sGRB 150101B are not suitable for macronova identification.  We also describe the HST observations of the afterglow emission of sGRB 050509B since they have not been formally reported in the literature yet.

\begin{itemize}
\item {GRB 050509B: this burst had dense HST followup observations in F814W band on May 14, 18, 28 and June 1 2005 (http://archive.stsci.edu), each exposure lasted 6908 s. There is a very bright elliptical galaxy ($\sim 16$ mag, $z=0.225$) near the {\it Swift} XRT error box, likely to be the host galaxy. Without an identification of an optical counterpart, it is challenging to set a robust limit on the underlying macronova (see also \cite{Hjorth2005} for similar conclusion on afterglow but based on the ground-based $R$/$V$ observations). If the GRB is  at the outskirt of the galaxy, the limit would be $>27.4$ mag and can be used to constrain the merger scenario. However if it is in the inner core of the galaxy, the limit would be $>25.3$ mag and the constraint on the macronova emission is weak.}

\item sGRB 050724: this burst had rare ground-based $I$-band observations and no HST followup observation, the large amount of dust extinction (i.e., $A_{\rm v}\approx 2~{\rm mag}$) \cite{Fox2005,Malesani2007b} in the direction of the burst is an important obstacle for macronova search, too.

\item {sGRB 060502B: the redshift of this burst is suggested to be either unknown \cite{Berger2010} or 0.287 \cite{Berger2014}. Assuming a redshift of 0.287, the main obstacle for macronova hunting is the lack of dense and deep optical/nIR observations and there was just a $R-$band flux upper limit $\leq 0.7\mu {\rm Jy}$ at $\sim 16.8$ hours reported \cite{Berger2010}. Considering its relatively high redshift, a macronova signal, if intrinsically as luminous as that identified in sGRB 050709, can just give rise to $R-$band peak emission of $\sim 26$th mag, which is well below the single upper limit previously reported.}

\item sGRB 071227: this event had a relatively high redshift ($z=0.381$) but had neither ground-based $I$-band observations nor HST followup observations \cite{Fong2015}.

\item sGRB 080905A: the redshift of $z=0.122$ is low that is suitable for macronova hunting. However, the latest two VLT/R-band followup observations were on Sept 7 and 23, respectively \cite{Fong2015}. Such rare observations, either too early or too late, are not helpful for macronova search (This is in particular the case for the NS-NS merger scenarios for which the peak R-band macronova emission is expected to be $\sim 26$th mag for $M_{\rm ej}\leq 0.01~M_{\odot}$ and $z\approx 0.1$ \cite{Tanaka2013,Kasen2013}.)

\item sGRB 140903A: this burst was at $z=0.351$, for which HST nIR observations are necessary to get the macronova signal. However, no HST exposure of the afterglow of GRB 140903A was performed,
the available dataset is not deep enough to search for a macronova (also discussed in Troja et al \cite{Levan2014}).

\item sGRB 150101B: the redshift is $z=0.134$.
If the associated kilonova emission is similar to that of GRB 130603B, a H-band peak magnitude is expected to be around ${\rm H(AB)=22-23}$ \cite{Levan2015a}, The IR observations with VLT/HAWK-I on January 16 however found no evidence
for any source to a preliminary limiting magnitude of H(AB)$>$23.5 \cite{Van2015}. The macronova signal of GRB 130603B was detected at $t\sim 7$ day (in the burster's rest frame), the non-detection in GRB 150101B in H-band may be due to the long delay of the exposure. The TNG had two J-band measurements on Jan 11, Jan 15 2015, respectively. The obtained upper limits, however, are not tight enough to exclude the presence of  a macronova as bright as that of GRB 130603B \cite{Fong2015}. The VLT I-band observations were performed either too early or too late for the macronova detection. The first visit of HST in F606W-band was on 11 Feb. 2015, which is about 40 days after the GRB trigger, too late to catch the macronova signal.
\end{itemize}

\begin{tiny}
\begin{table*}
\label{tab-Summary}
\begin{center}
\title{}Supplementary Table 1. The `nearby' sGRBs and hGRBs with optical afterglow emission.\\
\begin{tabular}{llllll}
\hline
	GRB	&	$z$		&	Ground-based I-band &	HST nIR observations & Macronova search\\
\hline
050509B$^{a}$	&	0.225 &	&	May 14, 18, 28 and Jun 1	& Suitable (No)	  	\\
 050709$^{a,b}$	&	0.161	&	VLT/Jul 12, 14, 18, 30	&	Jul 15, 19, 28 and later & Suitable (Yes)	 \\
 050724$^{a,c}$ 	&	0.257 &	VLT/Jul 25, 27, 30	&		 & Not Suitable	 \\
 060502B$^{d}$ 	&	0.287 	&		&		& Not Suitable \\
 060505$^{e}$ 	&	0.089 	&	Gemini-S/May 14	&	May 19 and Jun 06	& Suitable (No)	\\
 060614$^{f}$    &	0.125 	&	VLT/Jun 16, 17, 18 and later 	&	Jun 28 and Jul 16   & Suitable (Yes)  \\
 061201$^{a,g}$    &	0.111 	&	VLT/Dec 2, 3, 5, 18	&	 Dec. 11    & Suitable (No)   \\
 071227$^{a,h}$	&	0.381 	&		&		& Not Suitable \\
 080905A$^{a}$	&	0.122	&		&			& Not Suitable \\
 130603B$^{a,i}$	&	0.356 	&	Gemini/Jun 4, 5	&	Jun 13 and Jul 03		& Suitable (Yes) \\
 140903A$^{a,j}$	&	0.351 	&	Gemini-N/Sep 5	&		& Not Suitable	\\
 150101B$^{a,k}$   &	0.134 	& VLT/Jan 4, 19, 20		&	    & Not Suitable   \\
\hline
\end{tabular}
\end{center}

Note: a. Fong et al. \cite{Fong2015}; b. Fox et al. \cite{Fox2005} and Covino et al. \cite{Covino2005}; c. Malesani et al. \cite{Malesani2007b}; d. Berger et al. \cite{Berger2010}
e. Ofek et al. \cite{Ofek2007b}; f. Della Valle et al. \cite{Della2006b} and Gal-Yam et al.\cite{Gal-Yam2006b};
g. Stratta et al. \cite{Stratta2007b}; h. D'Avanzo et al. \cite{DAvanzo2009}; i. Tanvir et al. \cite{Tanvir2013}; j. Troja et al. \cite{Levan2014}; k. Levan et al. \cite{Levan2015a}.

\end{table*}
\end{tiny}

\end{appendix}

\end{document}